# Ricean Shadowed Statistical Characterization of Shallow Water Acoustic Channels for Wireless Communications

F. Ruiz-Vega, M. C. Clemente, P. Otero, and J. F. Paris

*Abstract*—In this letter, the statistical behaviour of the shallow water acoustic channel for wireless communications is shown to be well characterized by the Ricean shadowed distribution, which has never been proposed for communication purposes on this type of channel. This characterization is clearly motivated from statistical and physical perspectives and has both theoretical and practical advantages compared to previously proposed models.

## I. INTRODUCTION

ACOUSTIC channel characterization is essential in modem design for underwater wireless communications [1], [2]. So far, there is not a general agreement about the most convenient channel model [3]. In [4], three statistical distributions are proposed, Rice and Nakagami-*m* distributions being the best fitting ones. The model presented in [2] consists in Rayleigh short term variations, and a variable mean power that depends on physical environment parameters. Rice distribution is proposed in [1] by considering that every propagation path contains a dominant component together with many weak scattered components. Experimental results are reported in [5], where the K-distribution is shown to fit the shallow water acoustic channel variations using three different frequency resolutions. However, authors admit that this result is just a mathematical function fitting, and wonder about the underlying mechanism for signal fluctuation that leads to the K-distribution model. In this letter, it is shown that the Ricean shadowed distribution achieves characterization of the channel both in its statistical aspect and from a physical point of view. Although Ricean shadowed distribution has been already used in satellite communication models [6], it has not been proposed so far as a model of underwater acoustic communication channel.

In this paper, Ricean shadowed distribution is proposed for modelling underwater acoustic communication channel.

Section II states both the statistical and physical motivation of using Ricean shadowed distribution. In section III, statistical characterization is performed and some results are shown. Conclusion is presented in section IV.

## II. MOTIVATION

### A. Statistical Characterization

From statistical point of view, Ricean distribution arises from the addition of a Rayleigh distributed random component to a deterministic dominant component. The so-called Rice parameter *K* is defined as the ratio between the two components. The Ricean shadowed distribution can be obtained considering a Ricean distribution where the dominant component is also a random variable, which follows a Nakagami-*m* distribution with gamma power statistics [6]. Statistical motivation for considering Ricean shadowed is clear since Rayleigh and Rice distributions are particular cases of the Ricean shadowed distribution, being the latter a generalization of those.

### B. Physical Characterization

From a physical point of view, Rice and Rayleigh distributions have the well known interpretation of a multipath propagation scenario with or without a Line Of Sight (LOS) dominant component, and have been proposed so far by most authors to model the underwater acoustic channels [1]-[3]. Interestingly, the Ricean shadowed distribution provides a more comprehensive physical explanation of the propagation of acoustic signals over shallow water channels, since the dominant component changes from being deterministic to a random variable, modelling fluctuations. Note that acoustic wavelengths for frequencies over 5 kHz become fractions of meter, so they could be analyzed as underwater acoustic microwaves, enabling the use of geometric optics to model their propagation. As illustrated in Fig.1, the multipath components are due to multiple reflections in the sea surface and bottom, as well as bottom protuberances at the vicinity of the receiver, and the dominant component comes from line-of-sight (LOS) or strong reflection paths that are varying their amplitude because of water movement (due to ocean currents, pressure and temperature gradients, salinity variations) that causes scintillation of dominant LOS paths in Fig.1. This scintillating phenomenon is present as well in other propagation problems where the particles of a medium supporting an oscillatory perturbation are moving (even slower than sea water), producing instantaneous random variations of the refractive index [7] with consequent amplitude fluctuations (giving rise to the various close dominant LOS paths in Fig.1). Besides providing a physical interpretation of the underwater acoustic propagation underlying phenomena, the Ricean shadowed distribution enables performance analysis of a variety of wireless communication systems [8]-[10], unlike the K-distribution, the complexity of which renders it inconvenient for analytical performance evaluations, as noted in [10].

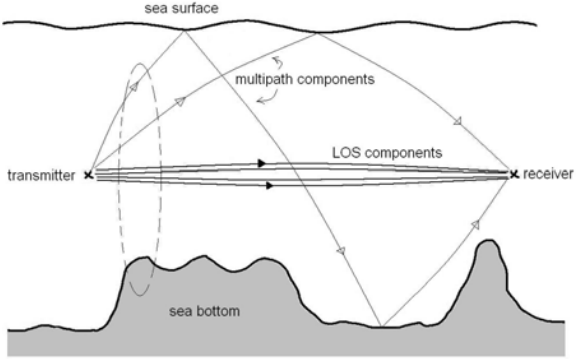

Fig. 1. Scenario with partial obstructed LOS component.

### III. STATISTICAL CHARACTERIZATION

A comparison on how the Ricean shadowed and the K-distributions probability density functions (PDF) model some shallow water acoustic channel measurements found in literature is presented below. The Ricean shadowed PDF of the channel response envelope $X$ is [10]

$$p_X(x) = \left(\frac{2b_0 m}{2b_0 m + \Omega}\right)^m \frac{x}{b_0} \exp\left(-\frac{x^2}{2b_0}\right) {}_1F_1\left(m,1;\frac{\Omega x^2}{2b_0(2b_0 m + \Omega)}\right) \quad (1)$$

where $\Omega$ is the average power of the LOS component, $2b_0$ is the average power of the scattered component, $m$ is a parameter ranging from 0 to ∞ that describes the severity of shadowing and $_1F_1$ is the Kummer confluent hypergeometric function [11]. According to above definitions, the Ricean parameter used in this letter is given by $K=\Omega/2b_0$. Regarding the experimental results published in reference [5], data were collected at sea after transmitting a signal centered at 17 kHz with 5 kHz bandwidth and used to calculate the PDF of the envelope fluctuations. The experiment was conducted by the Naval Research Laboratory, transmitting data from a fixed source to a fixed receiver at a distance of 3.4 km. Water depth in the experimental area was about 70 meters, and both source and receiver transducers were vertical arrays containing 8 elements at 35 meters depth. The process was performed using M-ary frequency-shift-keying consisting of many narrowband signals at frequencies separated by $\Delta f$ for various frequency resolutions ($\Delta f$ of 5, 80 and 320 Hz). Transmitted data size was 7102 M-ary frequency-shifted sequences. For the sake of fairness in the comparison between both distributions, the $\alpha$ scale parameter of the K-distribution ($\upsilon$ shape parameter is reported in [5]) has been optimized to fit the PDF of experimental data. Eventually, optimal values of the Ricean shadowed distribution, $K$ and $m$, were obtained to fit the experimental PDF. Parameters optimization was conducted using commercial optimization libraries. Specifically, a sequential quadratic programming method was used, in which the Hessian of the Lagrangian was calculated at each iteration [12]. In order to compare the goodness of fits, density functions are plotted in Fig. 2, for the 80 Hz case described above. The corresponding values of $K$ and $m$ (Ricean shadowed distribution), and $\upsilon$ and $\alpha$ (K-distribution), that yield the best fit are shown in the legends. The best fit is the one that yields the smallest mean squared error between both analytical PDF's and the experimental PDF. Moreover, the calculated mean squared errors are shown in Table I. It can be seen that the Ricean shadowed distribution fits the experimental data with practically the same accuracy as the K-distribution for $\Delta f$ =320 Hz and $\Delta f$ =5 Hz, and slightly better for $\Delta f$ =80 Hz.

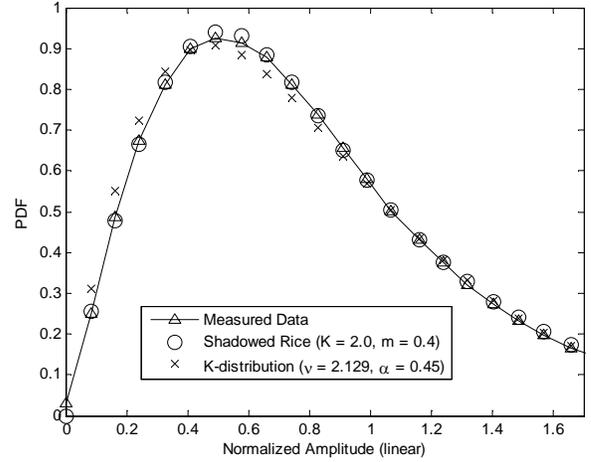

Fig. 2. Ricean shadowed and K-distribution PDFs fit for $\Delta f$ = 80 Hz.

TABLE I
MEAN SQUARED ERRORS OF THE TWO FITTING DISTRIBUTIONS

| Frequency resolution, $\Delta f$ (Hz) | K-distribution mean squared error | Ricean shadowed distribution mean squared error |
|---|---|---|
| 320 | $1.1273 \cdot 10^{-4}$ | $3.3848 \cdot 10^{-4}$ |
| 80 | $4.6910 \cdot 10^{-4}$ | $6.7276 \cdot 10^{-5}$ |
| 5 | $3.4044 \cdot 10^{-5}$ | $4.3709 \cdot 10^{-5}$ |

### IV. CONCLUSION

In this letter, the Ricean shadowed distribution is for the first time proposed to statistically model scintillating phenomenon of dominant component propagating over shallow water acoustic communication channels. In addition to providing such physical interpretation, statistical fitting with previously published experimental data is good and the mathematical manipulation of the distribution expression is feasible.